\documentclass[prl,twocolumn,floatfix,superscriptaddress,showpacs,amsmath,amssymb]{revtex4}%
\usepackage{bm}

\usepackage{graphicx}
\usepackage{dcolumn}
\usepackage{amsmath}
\usepackage{amssymb}
\usepackage{epsfig}
\usepackage{color}

\begin{document}

\title{Flow rule, self-channelization and levees in unconfined granular flows.}
\author{S. Deboeuf}
\author{E. Lajeunesse}
\affiliation{Institut de Physique du Globe de Paris, 4 place Jussieu,  75252 Paris cedex 05, France}
\author{O. Dauchot}
\affiliation{CEA Saclay/SPEC, URA2464, L'Orme des Merisiers, 91 191 Gif-sur-Yvette, France}
\author{B. Andreotti}
\affiliation{Laboratoire de Physique et M\'ecanique des Milieux H\'et\'erog\`enes, 10 rue Vauquelin, 75005 Paris, France}

\begin{abstract}
Unconfined granular flows along an inclined plane are investigated experimentally. During a long transient, the flow gets confined by quasi-static banks but still spreads laterally towards a well defined asymptotic state following a non-trivial process. Far enough from the banks a scaling for the depth averaged velocity is obtained, which extends the one obtained for homogeneous steady flows. Close to jamming it exhibits a crossover towards a non local rheology. We show that the levees, commonly observed along the sides of the deposit upon interruption of the flow, disappear for long flow durations. We demonstrate that the morphology of the deposit builds up {\it during} the flow, in the form of an underlying static layer, which can be deduced from surface velocity profiles, by imposing the same flow rule everywhere in the flow.
\end{abstract}

\maketitle
Geophysical granular flows such as pyroclastic flows or debris avalanches self-channelize, forming a coulee surrounded by static banks, until they come to arrest and form a deposit~\cite{I97,HLL96} with levee/channel morphology. Recent laboratory experiments~\cite{FT04} involving short time interruption of a localized flow of dry granular material have reproduced such deposit morphology and underlined the need for a deeper knowledge of the rheology. Flows down an incline provide a natural configuration for studying the rheometry of dense granular media. In the case of homogeneous flows, experiments~\cite{P99} have shown that there is a minimum thickness $h_{\rm stop}(\theta)$ below which no flow occurs and a maximum one $h_{\rm start}(\theta)$ above which static layers spontaneously destabilise. Later it has been shown~\cite{P99,SLG03,FT04,GDR04} that the depth averaged velocity $\bar u$is related to the thickness $h$ by the flow rule:
\begin{equation}
\frac{\bar u}{\sqrt{gh}} = \beta \frac{h}{h_{\rm stop}(\theta)} + \alpha, 
\label{flowrule1}
\end{equation}
where $g$ is the gravity, leaving unclear the dependance of the non-dimensional parameters $\alpha$ and $\beta$ on both the kind of grains and the covering of the incline. For a flow of glass beads down an incline covered with glued beads, $\alpha=0$ and the flow rule is consistent with a local rheology~\cite{GDR04,dCEPRC05}. This rheology has been validated in various flow configurations, including non-steady and non-uniform flows as well as flows on erodible ground~\cite{GDR04,JFP} and  was recently extended to three dimensions~\cite{JFP06}. However, it does not verify $\bar u(h_{\rm stop})=0$ and hence does not describe the flow arrest, where non local effects are expected to become significant~\cite{Ra03}. Requiring this last condition imposes $\alpha=-\beta$, a situation actually reported in the case of unconfined flows~\cite{FT04}. Hence, it is of primary importance, both for practical and fundamental reasons to investigate further the situation where flowing and static regions coexist. 

In this letter, we study experimentally an {\it unconfined} flow down an incline. The flow indeed self-channelizes within static banks which may evolve freely. Our main goals are to characterise the rheology and dynamics of the flow close to jamming, and its relation to the morphology of the deposit when the flow stops.
\begin{figure}[t!]
\includegraphics{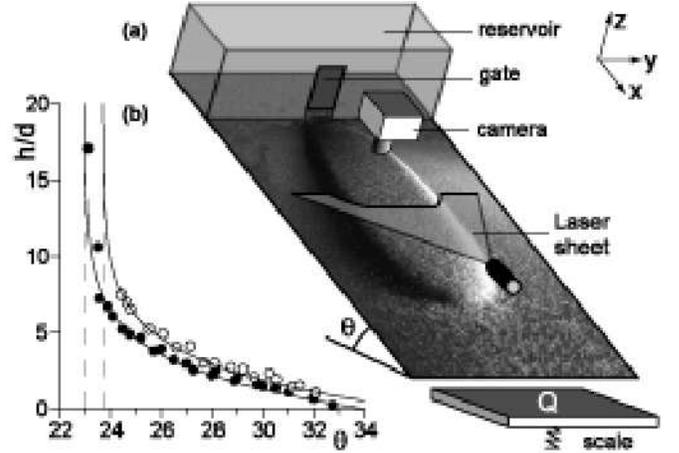}
\vspace{-5mm}
\caption{{\bf (a)} Sketch of the experimental setup. {\bf (b)} Phase diagram in the plane $(h,\theta)$: the experimental data for $h_{\rm stop}(\theta)$ ($\bullet$) and $h_{\rm start}(\theta)$ ($\circ$) with the best fits by eq.~(\ref{mumu}) (solid line).}
\vspace{-5 mm}
\label{setup}
\end{figure}

{\it Set-up --} The experimental setup is sketched on Fig.~\ref{setup}(a). A $60 \times 300~{\rm cm^2}$ plane, covered with sand paper of average roughness about  $200~{\rm \mu m}$, is inclined at an angle $\theta$ ranging from $24^\circ$ to $34^\circ$. The granular material consists of a slightly polydisperse mixture of spherical glass beads of diameter $d=350\pm50~\rm{\mu m}$ and density $\rho=2500~{\rm kg.m}^{-3}$. Grains are released from a reservoir located at the top of the plane by opening  a gate of adjustable height and width. To produce homogeneous flows, the whole plane width is used, but for unconfined flows the gate width is reduced to $5$~cm and the width of the plane is always larger than that of the avalanche. For the present set up, $h_{\rm start}(\theta)$ and $h_{\rm stop}(\theta)$ obey (Fig.~\ref{setup}b):
\begin{equation}
\tan \theta = \tan \theta^\infty+\left(\tan \theta^0-\tan \theta^\infty\right)\exp \left(-h/\alpha d\right) , 
\label{mumu}
\end{equation}
with $\alpha=3.2$, $\theta^\infty_{\rm stop}=22.5^{\circ}$, $\theta^\infty_{\rm start}=23.0^{\circ}$, $\theta^0_{\rm stop}=33.6^{\circ}$, and $\theta^0_{\rm start}=35.6^{\circ}$. The mass flow rate $Q$  is measured at the end of the plane, where the grains fall inside a reservoir resting on a scale. During all experiments, some of which lasted up to $5$~hours, $Q$ fluctuated by less than $2\%$. A 572 $\times$ 768 pixels camera positioned at the vertical of the plane is used to acquire images of the flow at $25$~Hz. The local flow thickness $h$ is measured using the deviation of a laser sheet inclined at $\sim 5^{\circ}$ over the layer. The measured sensitivity ($\sim 60~\rm{\mu m}$) is smaller than the grain size thanks to the averaging across the beam width $(10 {\rm mm})$ along $x$. About $10\%$ of the grains are dyed in black and the velocity  of the surface grains, $u_s$, is measured using a particle-imaging velocimetry algorithm. In all experiments the spanwise velocity component is found to be smaller than the resolution ($2$~mm/s).
\begin{figure}[t!]
\includegraphics{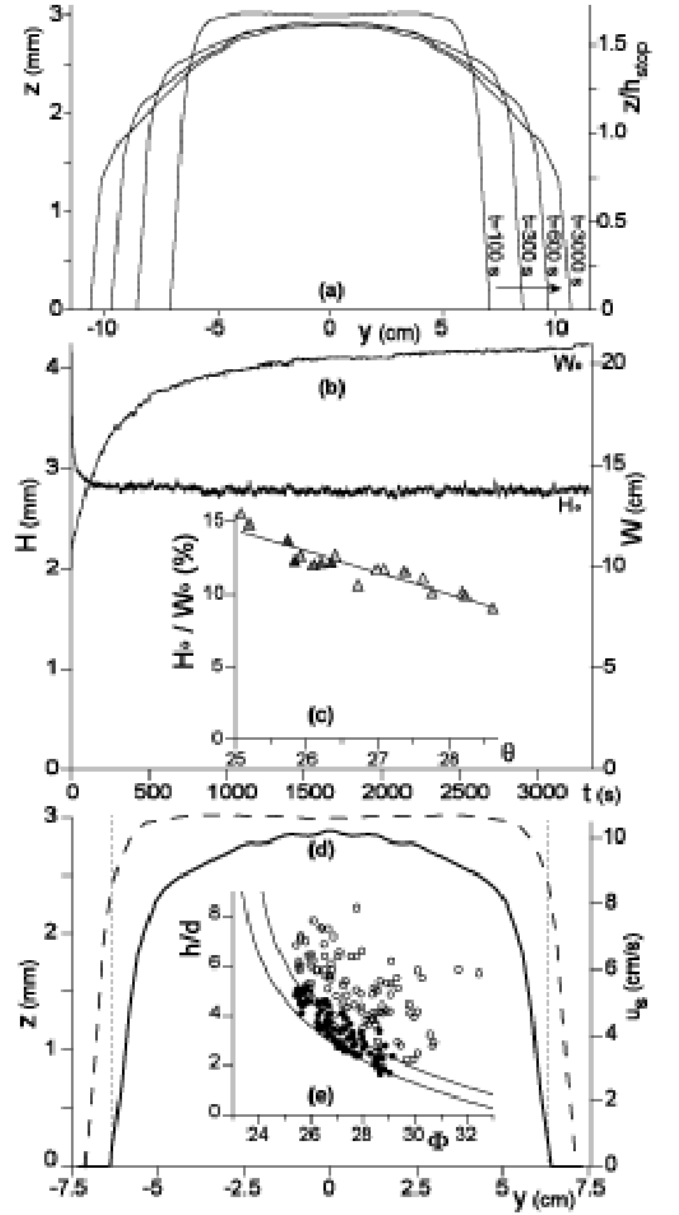}
\vspace{-5mm}
\caption{{\bf (a)} Time evolution of the height profiles $h(y,t)$  and {\bf (b)} evolution of the central height $H$ and width $W$ of the flow for $Q=25~{\rm g.s}^{-1}$, $\theta=25^\circ$. {\bf (c)} Asymptotic aspect ratio of the flow  $H_{\infty}/W_{\infty}$ as a function of the slope angle, for flow rates $Q$ ranging from $3~$g/s ($\triangle$) to $30~$g/s ($\blacktriangle$). {\bf (d)} Velocity $u_s(y)$ (solid line) and thickness $h(y)$ (dashed line) profiles for $\theta=25^{\circ}$, $Q=25~{\rm g.s}^{-1}$ and $t=150$~s. {\bf (e)} Thickness vs. local slope $\Phi$ inside the quasi-static banks $(\circ ): t<2000$~s, $(\bullet )$: asymptotic state. The solid lines show  $h_{\rm stop}$ and $h_{\rm start}$. }
\vspace{-5mm}
\label{morpho}
\end{figure}

{\it Transverse spreading \& self-channelization--} Upon release of the grains, an avalanche front propagates down the plane at a constant velocity, leaving behind it a streamwise flow uniform in the x-direction (Fig.~\ref{setup}a). Fig.~\ref{morpho}(a) displays thickness profiles at successive time steps, after the front has reached the downstream extremity of the incline -- typically after $100~$s. It shows that the flow widens while the thickness profiles bend progressively. The central flow thickness converges rapidly toward an asymptotic value $H_{\infty}(\theta,Q)$, whereas its width converges to an asymptotic value $W_{\infty}(\theta,Q)$ with a much larger relaxation time (Fig.~\ref{morpho}b). When preparing a flow of width larger than $W_{\infty}$ by increasing the flow rate, and returning to the initial flow rate, the width decreases back to $W_{\infty}$. This shows beyond any doubt the selection of an asymptotic steady state. Both $W_{\infty}$ and $H_{\infty}$ are observed to increase with the flow rate $Q$. The aspect ratio $H_{\infty}/W_{\infty}$ turns out to be independent of $Q$ and slightly decreases with $\theta$ (Fig.~\ref{morpho}c).

Fig.~\ref{morpho}(d) shows the transverse structure of the flow: an inner flowing region ($u\neq0$) flanked by two static banks ($u=0$) of width approximately $5~{\rm mm}$. These banks forms due to a strong increase of friction on the side of the layer, as the thickness $h$ vanishes and thus $\mu$ tends to $\mu_0=\tan \theta_0$ (eq.~(\ref{mumu})). Once the banks are formed, their external sides are so steep that the free surface angle $\tan\Phi \equiv (\tan^2(\theta)+(\partial_y h)^2)^{1/2}$ becomes significantly larger than the plane angle $\theta$ (Fig.~\ref{morpho}e) allowing transverse displacements. Indeed, the banks are outside the metastable band  $[h_{\rm stop},h_{\rm start}]$ during the spreading phase and converge toward it in the asymptotic state. The flow is hence divided into a central flow region uniform in the y-direction --~the ratio of the transverse to the vertical diffusion of momentum $h^2/W^2 \approx 10^{-4}$~-- and the banks dominated by three-dimensional effects. In both regions, inertial terms $ u_y \partial u_x/ \partial y \approx h u_s^2/W^2 \approx 10^{-3} g $ are small.
\begin{figure}[t!]
\includegraphics{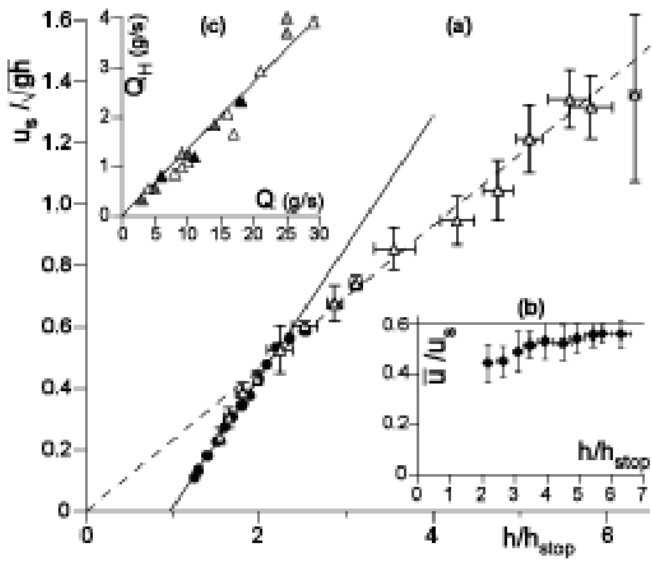}
\vspace{-5mm}
\caption{{\bf (a)} Flow rule $u_s/(gh)^{1/2}$ vs $h/h_{\rm stop}$: $(\bigtriangleup )$ data for steady homogeneous flows, $(\bullet)$ data obtained in the central --~uniform~-- part of unconfined flows. Each point is an average over different realizations and the corresponding statistical error bars are reported. The solid (resp. dashed) line is the best fit by eq.~(\ref{flowrule2}) (resp. eq.~(\ref{flowrule1}) with $\alpha=0$). {\bf (b)} Average to surface velocity ratio $\bar u/u_s$ vs. $h/h_{\rm stop}$ for steady homogeneous flows. {\bf (c)} Scaling of the asymptotic height with the flow rate:  $ Q_H = \rho g^{1/2} H_{\infty}^{5/2} (H_{\infty}/h_{\rm stop}-1)$ as a function of $Q$, for $\theta$ ranging from $25^\circ$ ($\triangle$) to $28^\circ$ ($\blacktriangle$).}
\vspace{-5mm}
\label{flowrule}
\end{figure}

{\it Flow rule --} The flow rule is first determined using homogeneous steady flows covering the total width of the plane (Fig.~\ref{flowrule}a, $\bigtriangleup$). For $h/h_{\rm stop}>2$ it is well described by eq.~(\ref{flowrule1}) with $\alpha=0$ and $\beta=0.134$ assuming $\bar u/u_s=3/5$ (see below). For $h/h_{\rm stop}<2$, there is a systematic deviation towards lower $u_s/(gh)^{1/2}$ values. However smaller thicknesses can not be further investigated in homogeneous flows. On the contrary self-channelized flows are naturally very thin and give access precisely to that range of thickness. Fig.~\ref{flowrule}a ($\bullet$) shows data collected near the centerline of the flow, where the flow has been shown above to be uniform in the transverse direction. Plotting  $u_s/\sqrt{g h}$  as a function of $h/h_{\rm stop}$, one observes again a collapse of data (for different $\theta$, $Q$ and $t$) on a single curve which satisfies $\bar u(h_{\rm stop})=0$ and can be described at the first order in $h/h_{\rm stop}$ by:
\begin{equation}
\frac{\bar u}{\sqrt{gh}} = \tilde\beta \left(\frac{h}{h_{\rm stop}(\theta)} - 1\right), 
\label{flowrule2}
\end{equation}
with $\tilde\beta=0.219$ assuming now $\bar u/u_s=1/2$ (see below). Note that both in the confined and unconfined cases, we have checked using dyed grains that the flow involves the whole layer thickness. Most importantly, the data obtained in both cases coincide on their common range of thickness and provide -- to the best of our knowledge -- the first experimental determination of the flow rule over the whole range of thicknesses, independently of the flow configuration.

We now briefly discuss its relation to the locality of the rheology.  For a large enough thickness, the flow rule obeys eq.~(\ref{flowrule1}) with $\alpha=0$ and is consistent with a local rheology~\cite{GDR04}~: when the shear stress $\tau$ depends on the shear rate $\dot \gamma$ at the considered point only, it follows from dimensional analysis -- assuming a small transverse shear stress as suggested by the flatness of the transverse velocity profiles -- that the rheology can be expressed under the non-dimensional form $\tau/P=\mu_I(I)$ with $I=\dot\gamma d/(P/\rho)^{1/2}$, where $d$ is the grain diameter, $P$ the pressure and $\rho$ the density. Integrating this last relation leads to a scaling law for the depth averaged velocity $\bar u(h,\theta) = A(\theta) h^{3/2}$, which is indeed satisfied by eq~(\ref{flowrule1}), when $\alpha=0$. For a small thickness the flow rule eq.~(\ref{flowrule2}) clearly displays a violation of the local rheology close to jamming, which is confirmed by the measurement of the ratio of the depth averaged velocity $\bar u$ --~measured with the avalanche front velocity in the homogeneous flow case~-- to the surface velocity $u_s$ (Fig.~\ref{flowrule}b). For large $h/h_{\rm stop}$ values, $\bar u/u_s=3/5$ which is consistent with Bagnold-like profile deriving from a local rheology~\cite{GDR04}. Close to jamming transition, $\bar u/u_s$ decreases to $1/2$  in agreement with numerical findings~\cite{SLG03} of a transition toward a linear velocity profile. Note that the above analysis suggests that $\beta/h_{\rm stop}(\theta)$ shall not depend on the covering of the incline whereas  $\tilde\beta/h_{\rm stop}(\theta)$ and of course $h_{\rm stop}$ could.
\begin{figure}[b!]
\includegraphics{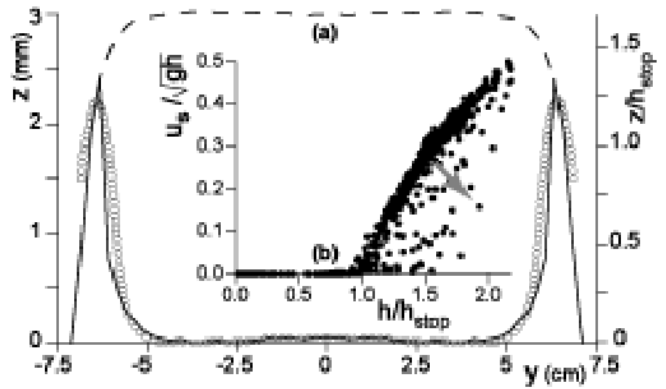}
\vspace{-5mm}
\caption{{\bf (a)} Static layer thickness $Z$ reconstructed from surface velocity measurements (solid line) and directly measured with a brush ($\circ$) -- dashed line : total height. {\bf (b)} $u_s/(gh)^{1/2}$ vs $h/h_{\rm stop}$, data obtained across the whole flow width: there is a progressive shift from the flow rule when moving away from the centerline (arrow).}
\vspace{-3mm}
\label{banks}
\end{figure}
Finally from conservation of mass, one expects a linear relationship between $Q$ and $H_\infty W_\infty \bar u(H_\infty) \propto \rho g^{1/2} H_\infty^{5/2} (H_\infty/h_{\rm stop}-1)$, using eq.~(\ref{flowrule2}). This relation is  closely verified experimentally (Fig.~\ref{flowrule}c) and suggests to describe the flow in a shallow water approximation. However, let us mention that neither the asymptotic state (its bended shape but also the selection of the aspect ratio $H_\infty/W_\infty$) nor the temporal scalings of $W(t)$ and $H(t)$ can be {\it simply} predicted following such an approach. Hence we shall postpone this analysis as well as a full three-dimensional analysis to a forthcoming paper. Note that the complete set of shallow water equations is solved in~\cite{MVBT04}.

{\it Banks~--} In the previous section, we investigated the flow rule near the centerline of the flow. No such scaling is {\it a priori} expected close to the static banks. Indeed as seen on Fig.~\ref{banks}(b), when acquiring data moving away from the centerline, no scaling is observed and $u_s/(gh)^{1/2}$ is systematically underestimated. A possible explanation of such a deviation from the otherwise scaling is the extension of the static zone below the flowing grains. In order to verify such an hypothesis, we have performed specific experiments.

One first conducts an experimental run using white grains. After a few minutes, the flow is stopped and a deposit forms. The white grains of a small slice across the deposit are removed and replaced by black grains of the same material. The flow is then started again at the same flow rate. At  the  surface, the  black grains are washed out by the incoming white grains except in the static borders. After a few minutes, the flow is interrupted again. Using a brush, the white grains are removed very cautiously layer after layer in the region of the slice. Close to the centre, the deposit is exclusively composed of the new white grains, proving that the flow involves the whole thickness. However, close to the banks, a layer of black grains remained trapped, indicating the presence of a static~\cite{rmq} layer of thickness $Z(y,t)$ (denoted by $\circ$ in Fig.~\ref{banks}a) below the flowing one. The interface between black and white grains turns out to be very sharp (of the order of a single grain diameter). Using $u_s$, one can construct the thickness $R \leq  h$ which would flow if the flow rule obtained above were applicable inside the flowing layer. The reconstructed static/flowing interface $Z=h-R$ is shown on Fig.~\ref{banks}(a) together with the measured one. The fairly good agreement suggests that, to the first order, mobile grains flow above  static grains just like they flow above the rough plane. At the same order, it tells us that the flow rule is actually observed everywhere in the flow when applied to the flowing thickness $R$ instead of the total one $h$.

\begin{figure}[t!]
\includegraphics{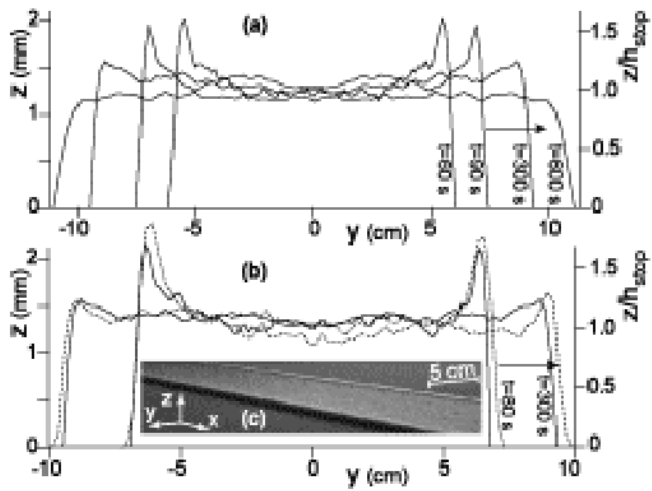}
\vspace{-5mm}
\caption{(a) Thickness profiles of the deposit $h_{\rm dep}(y)$ obtained after different flow durations for $Q=25~{\rm g.s}^{-1}, \theta=26^{\circ}$. (b) Comparison between measured (solid line) and predicted (dotted line) deposit profiles. (c) Picture of levees observed at the borders of a deposit. }
\vspace{-5mm}
\label{Deposit}
\end{figure}

{\it Levees~--} Turn now to the morphology of the deposit which forms when the flow stops. It has been reported~\cite{FT04} that, under certain conditions of inclination and flow rate, the deposit formed upon interruption of the flow exhibits a levee/channel morphology similar to those observed on pyroclastic flow deposits.  In the present study, we have observed that for larger times, the flow actually keeps on widening and converges only very slowly toward its asymptotic state. Accordingly the  shape of the deposit strongly depends on the flow duration $t$ (Fig.~\ref{Deposit}a). For small $t$,  deposits are composed  of a central flat zone of thickness $h_{\rm stop}$ bordered by two levees of thickness larger than $h_{\rm stop}$ (Fig.~\ref{Deposit}c) as previously reported \cite{FT04}. When $t$ increases, the levee thickness  decreases until it vanishes at very long time, so that the deposit corresponding to the asymptotic state is indeed flat. Levees result from the combination between lateral static zones on each border of the flow and the drainage of the central part of the flow after the supply stops~\cite{FT04}. However, a clear picture is still lacking concerning the junction between a central flat zone of thickness $h_{\rm stop}$ and a levee of thickness larger than $h_{\rm stop}$. The flow rule that was obtained here provides a very simple scenario.  The flow stops when $R=h_{\rm stop}$, and one expects the deposit to form by superimposing a layer of thickness $h_{\rm stop}$ to the static layer. Accordingly the deposit thickness should be $h_{\rm dep}=Z+h_{\rm stop}$ wherever $u_s \neq 0$ and $h_{\rm dep}=Z$ in the quasi-static banks, where $u_s=0$. Fig~\ref{Deposit}(b) provides the experimental evidence that such a simple scenario indeed holds: the predicted and the measured deposit profiles qualitatively match.

To conclude, by investigating unconfined granular flows down an incline, we have shown that they obey a flow rule which reveals a crossover towards a non local rheology close to the jamming transition. This flow rule accounts for the morphology of the deposit, which actually builds up {\it during} the flow, in the form of an underlying static layer. This underlines the importance of addressing erosion-deposition mechanisms issues to complete a full description of geophysical flows.

The authors acknowledge E. Cl\' ement, F. Malloggi, J-P. Vilotte for fruitful discussions, Y. Gamblin, A. Viera, G. Bienfait, G. Simon for technical assistance and the student L. Baures for his work.


\begin{thebibliography}{99}
\bibitem{rmq} The velocity ($<10^{-6}~$m/s) is at least five orders of magnitude smaller than that in the flowing zone.
\end{thebibliography}
\end{document}